\documentclass[twocolumn]{el-author}
\usepackage{algorithmic}

\newcommand{\ua}{\uparrow}
\newcommand{\nc}{\newcommand}
\nc{\da}{\downarrow} \nc{\hc}{\hat{c}} \nc{\hS}{\hat{S}}
\nc{\bra}{\langle} \nc{\ket}{\rangle} \nc{\eq}{equation (\ref}
\nc{\h}{\hat} \nc{\hT}{\h{T}}\nc{\be}{\begin{eqnarray}}
\nc{\ee}{\end{eqnarray}}\nc{\rd}{\textrm{d}}
\nc{\hR}{\hat{R}}\nc{\Tr}{\mathrm{Tr}}
\nc{\tS}{\tilde{S}}\nc{\tr}{\mathrm{tr}}\nc{\8}{\infty}\nc{\lgs}{\bra\ua,\phi|}\nc{\rgs}{|\ua,\phi\ket}
\nc{\hU}{\hat{U}}\nc{\lfs}{\bra\phi|}\nc{\rfs}{|\phi\ket}\nc{\hZ}{\hat{Z}}\nc{\hd}{\hat{d}}\nc{\mD}{\mathcal{D}}
\nc{\bd}{\bar{d}}\nc{\bc}{\bar{c}}\nc{\mc}{\mathcal}\nc{\ea}{eqnarray}\nc{\mG}{\mathcal{G}}\nc{\bce}{\begin{center}}
\nc{\ece}{\end{center}}

\newcommand{\A}{\boldsymbol{A}}
\newcommand{\B}{\boldsymbol{B}}
\newcommand{\C}{\boldsymbol{C}}
\newcommand{\I}{\boldsymbol{I}}
\newcommand{\y}{\boldsymbol{y}}
\newcommand{\x}{\boldsymbol{x}}
\newcommand{\1}{\boldsymbol{1}}
\newcommand{\e}{\boldsymbol{e}}

\begin{document}

\title{Improved Bounds on the Restricted Isometry Constant for Orthogonal Matching Pursuit}

\author{Jinming Wen, Xiaomei Zhu and Dongfang Li}

\abstract{In this letter, we first construct a counter example to show that for any given positive integer $K\geq 2$ and for any $\frac{1}{\sqrt{K+1}}\leq t<1$, there always exist a $K-$sparse $\x$ and a matrix $\A$ with  the restricted isometry constant $\delta_{K+1}=t$ such that the OMP algorithm fails in $K$ iterations. Secondly, we show that even when $\delta_{K+1}=\frac{1}{\sqrt{K}+1}$, the OMP algorithm can also perfectly recover every $K-$sparse vector $\x$ from $\y=\A\x$ in $K$ iteration. This improves the best existing results which were independently given by Mo  et al. and Wang  et al.
}

\maketitle

\section{Introduction}
Consider the following linear model:
\begin{equation}
\label{e:model}
\y=\A\x
\end{equation}
where $\x\in \mathbb{R}^n$ is an unknown signal, $\y\in \mathbb{R}^m$ is an observation vector and $\A\in \mathbb{R}^{m\times n}$ (with $m<<n$) is a known sensing matrix. This model arises from compressed sensing, see, e.g., \cite{Don06} and one of the central goals is to recover $\x$ based on $\A$ and $\y$. It has been shown that under some suitable conditions, $\x$ can be recovered exactly, see, e.g., \cite{CanT05}.

The orthogonal matching pursuit (OMP) \cite{TroG07} is one of the commonly used algorithms to recover $\x$ from \eqref{e:model}. A vector $\x\in \mathbb{R}^n$ is $k-$sparse if $|\text{supp}(\x)|\leq k$, where $\text{supp}(\x)=\{i:x_i\neq0\}$ is the support of $\x$.
For any set $T\subset\{1,2,\ldots,n\}$, let $\A_T$ be the submatrix of $\A$ that only contains columns indexed by $T$ and
$\x_T$ be the restriction of the vector $\x$ to the elements indexed by $T$. Then the OMP can be described by Algorithm \ref{a:OMP}.

One of the commonly used frameworks for sparse recovery is the restricted isometry property,
which was introduced in \cite{CanT05}. For any $m\times n$ matrix $\A$ and any integer $k, 1\leq k\leq n$, the $k-$restricted isometry constant $\delta_k$ is defined as the smallest
constant such that
\begin{equation}
\label{e:RIP}
(1-\delta_k)\|\x\|_2^2\leq \|\A\x\|_2^2\leq(1+\delta_k)\|\x\|_2^2
\end{equation}
for all $k-$sparse vector $\x$.

It has conjectured in \cite{DaiM09} that there exist a matrix with $\delta_{K+1}\leq\frac{1}{\sqrt{K}}$ and a $K-$sparse $\x$ such that the OMP fails in K iterations \cite{MoS12}. Counter examples have independently given in \cite{MoS12} and \cite{WanS12} that there exist a matrix with $\delta_{K+1}=\frac{1}{\sqrt{K}}$ and a $K-$sparse $\x$ such that the OMP fails in K iterations. In this letter, we will give a counter example to show that for any given positive integer $K\geq 2$ and for any $\frac{1}{\sqrt{K+1}}\leq t<1$, there always exist a $K-$sparse $\x$ and a matrix $\A$ with $\delta_{K+1}=t$ such that the OMP algorithm fails in $K$ iterations. This result not only greatly improves the existing results, but also gives a counter example with $\delta_{K+1}<\frac{1}{\sqrt{K}}$ such that the OMP fails in K iterations.

It has respectively shown in \cite{DavW10} and \cite{LiuT10} that $\delta_{K+1}<\frac{1}{3\sqrt{K}}$ and $\delta_{K+1}<\frac{1}{(1+\sqrt{2})\sqrt{K}}$ are sufficient for OMP to recover every K-sparse $\x$ in K iteration. The sufficient condition has independently improved to $\delta_{K+1}<\frac{1}{1+\sqrt{K}}$ in \cite{MoS12} and \cite{WanS12}. In this letter, we will improve it to $\delta_{K+1}\leq\frac{1}{1+\sqrt{K}}$.

\begin{algorithm}[h!]
\caption{OMP \cite{TroG07}, \cite{WanS12} }  \label{a:OMP}
Input: measurements $\y$, sensing matrix $\A$
 and sparsity $K$.\\
Initialize: $k=0, r^0=\y, T^0=\emptyset$.\\
While $k<K$
\begin{algorithmic}[1]
\STATE $k=k+1$,
\STATE $t^k=\arg \max_j|\langle r^{k-1},\A_j\rangle|$,
\STATE $T^k=T^{k-1}\bigcup\{t^k\}$,
\STATE $\hat{\x}_{T^k}=\arg \min_{\x}\parallel\y-\A_{T^k}\x\parallel_2$,
\STATE $r^k=\y-\A_{T^k}\hat{\x}_{T^k}$.
\end{algorithmic}
Output: $\hat{\x}=\arg \min_{\x: \text{supp}(\x)=T^K}\parallel\y-\A\x\parallel_2$.
\end{algorithm}

\section{Main Results} In this section, we will give our main results. We will first construct a counter example to show the OMP algorithm may fail in $K$ iterations if $\frac{1}{\sqrt{K+1}}\leq\delta_{K+1}<1$.

\begin{theorem} \label{t:count}
For any given positive integer $K\geq 2$ and for any
\begin{equation*}
 \frac{1}{\sqrt{K+1}}\leq t<1
\end{equation*}
there always exist a $K-$sparse $\x$ and a matrix $\A$ with  the restricted isometry constant
$\delta_{K+1}=t$ such that the OMP  fails in $K$ iterations.
\end{theorem}

Our proof is similar to the method used in \cite{MoS12}, but the critical idea is different.

{\bf Proof.}  For any given positive integer $K\geq 2$, let
\begin{equation*}
\B=\begin{bmatrix}
\frac{K}{K+1}\I_{K}&  \frac{\1}{K+1} \\ \frac{\1^T}{K+1} & \frac{K+2}{K+1}
\end{bmatrix}
\end{equation*}
where $\1$ is a $K-$dimensional column vector with all of its entries being $1$ and $\I_{K}$ is the $K-$dimensional identity matrix.

By some simple calculations, we can show that the eigenvalues $\{\lambda_i\}_{i=1}^{K+1}$ of $\B$ are
\begin{align*}
&\lambda_1=\ldots=\lambda_{K-1}=\frac{K}{K+1}, \lambda_{K}=1-\frac{1}{\sqrt{K+1}}, \; \lambda_{K+1}=1+\frac{1}{\sqrt{K+1}}.
\end{align*}
let $s=t-\frac{1}{\sqrt{K+1}}$ and $\C=\B-s\I_{K+1}$. Then by the aforementioned two equations, the eigenvalues $\{\lambda_i\}_{i=1}^{K+1}$ of $\C$ are
\begin{align*}
&\lambda_1=\ldots=\lambda_{K-1}=\frac{K}{K+1}-s \\
&\lambda_{K}=1-\frac{1}{\sqrt{K+1}}-s=1-t, \; \lambda_{K+1}=1+\frac{1}{\sqrt{K+1}}-s.
\end{align*}

Since $ \frac{1}{\sqrt{K+1}}\leq t<1$, $\C$ is a symmetric positive definite matrix. Therefore, there exists an upper triangular
matrix $\A$ such that $\A^T\A=\C$. By the aforementioned inequations and \eqref{e:RIP}, $\delta_{K+1}(\A)=t$.

Let $\x=(1,1,\ldots,1,0)\in \mathbb{R}^{K+1}$, then $\x$ is $K-$sparse. Let $\e_{i}, 1\leq i\leq K+1$, denote the $i-th$ column of $\I_{K+1}$, then
one can easily show that,
\begin{align*}
\frac{K}{K+1}-s=\max_{1\leq i\leq K}|\langle\A\e_i, \A\x\rangle|\leq |\langle\A\e_{K+1}, \A\x\rangle|=\frac{K}{K+1}
\end{align*}
so the OMP fails in the first iteration. Therefore, the OMP algorithm fails in $K$ iterations for the given vector $\x$ and the given matrix $\A$.

In the following, we will improve the sufficient condition $\delta_{K+1}<\frac{1}{\sqrt{K}+1}$ \cite{MoS12}, \cite{WanS12} of the perfect recovery
 to $\delta_{K+1}\leq\frac{1}{\sqrt{K}+1}$.

\begin{theorem} \label{t:main}
Suppose that $\A$ satisfies the restricted isometry property of order $K+1$ with the
restricted isometry constant
\begin{equation}
\label{e:cond1}
\delta_{K+1}=\frac{1}{\sqrt{K}+1}
\end{equation}
then the OMP algorithm can perfectly recover any $K-$sparse signal $\x$ from
$\y=\A\x$ in $K$ iteration.
\end{theorem}

Before proving this theorem, we need to introduce the following two lemmas, where Lemma \ref{l:main2} was proposed in \cite{WanS12}.
\begin{lemma} \label{l:main1}
For each $\x, \x'$ supported on disjoint subsets $S, S'\subseteq \{1,\ldots, n\}$ with $|S|\leq s, |S'|\leq s'$, we have
\begin{equation}
\label{e:eq1}
|\langle \A\x, \A\x'\rangle|=\delta_{s+s'}\|\x\|_2\|\x'\|_2
\end{equation}
if and only if:
\begin{equation}
\label{e:eq2}
\frac{\|\A\x\|_2^2}{\|\x\|_2^2}+\frac{\|\A\x'\|_2^2}{\|\x'\|_2^2}=2.
\end{equation}
\end{lemma}

{\bf Proof.} Let
\begin{equation}
\label{e:x}
\bar{\x}=\x/\|\x\|_2, \bar{\x}'=\x'/\|\x'\|_2
\end{equation}
since $S\bigcap S'=\emptyset$, we have,
$\|\bar{\x}+\bar{\x}'\|_2^2=\|\bar{\x}-\bar{\x}'\|_2^2=2$. By \eqref{e:RIP}, we have
\begin{equation}
\label{e:rip2}
2(1-\delta_{s+s'})\leq \|\A(\bar{\x}\pm\bar{\x}')\|_2^2\leq 2(1+\delta_{s+s'})
\end{equation}
By the parallelogram identity and \eqref{e:x}, we have
\begin{equation}
\label{e:par}
|\langle \A\bar{\x}, \A\bar{\x}'\rangle|=\frac{1}{4}|\|\A(\bar{\x}+\bar{\x}')\|_2^2-\|\A(\bar{\x}-\bar{\x}')\|_2^2|\leq \delta_{s+s'}.
\end{equation}
By \eqref{e:x},  \eqref{e:eq1} holds if and only if the equality in \eqref{e:par} holds. By \eqref{e:rip2}, the equality in \eqref{e:par} holds if and only if
\begin{equation*}
\label{e:RIP2}
\|\A(\bar{\x}+\bar{\x}')\|_2^2=2(1-\delta_{s+s'}), \|\A(\bar{\x}-\bar{\x}')\|_2^2= 2(1+\delta_{s+s'})
\end{equation*}
or
\begin{equation*}
\label{e:RIP2}
\|\A(\bar{\x}-\bar{\x}')\|_2^2=2(1-\delta_{s+s'}), \|\A(\bar{\x}+\bar{\x}')\|_2^2= 2(1+\delta_{s+s'}).
\end{equation*}
Therefore, \eqref{e:eq1} holds if and only if
\begin{equation*}
\label{e:RIP2}
\|\A(\bar{\x}-\bar{\x}')\|_2^2+\|\A(\bar{\x}+\bar{\x}')\|_2^2= 4.
\end{equation*}
Obviously, the aforementioned equation is equivalent to \eqref{e:eq2}, so the lemma holds.

\begin{lemma} \label{l:main2}
For $S\subset\{1,2,\ldots,n\}$, if $\delta_{|S|}<1$, then
\begin{equation*}
(1-\delta_{|S|})\parallel\x\parallel_2\leq\parallel\A_S^T\A_S\x\parallel_2\leq(1+\delta_{|S|})\parallel\x\parallel_2
\end{equation*}
for any vector $\x$ supported on $S$.
\end{lemma}

We will prove it by induction. Our proof is similar to the method used in \cite{MoS12}, but the critical idea is different.

{\bf Proof of Theorem \ref{t:main}} Firstly, we prove that if \eqref{e:cond1} holds, then the OMP can choose a correct
index in the first iteration.

Let $S$ denote the support of the $K-$sparse signal $\x$ and let $\alpha=\max_{i\in S}|\langle\A\e_i, \A\x\rangle|$.
Then
\begin{equation*}
|\langle\A\x,\A\x\rangle|=|\sum_{i\in S}x_i\langle\A\e_i,\A\x\rangle|\leq\alpha\parallel\x\parallel_1\leq\alpha\sqrt{K}\parallel\x\parallel_2.
\end{equation*}
By \eqref{e:RIP}, it holds that
\begin{equation}
\label{e:lb1}
|\langle\A\x,\A\x\rangle|\geq(1-\delta_{K+1})\parallel\x\parallel_2^2.
\end{equation}
By the aforementioned two inequations, we have
\begin{equation}
\label{e:lb2}
\max_{i\in S}|\langle\A\e_i, \A\x\rangle|\geq\frac{(1-\delta_{K+1})\parallel\x\parallel_2}{\sqrt{K}}
\end{equation}
and if the equality in \eqref{e:lb2} holds, then the equality in \eqref{e:lb1} must also hold.

By Lemma 2.1 in \cite{Can08}, for each $j\not\in S$, it holds
\begin{equation}
\label{e:ub}
|\langle\A\e_j,\A\x\rangle|\leq \delta_{K+1}\parallel\x\parallel_2.
\end{equation}
So if \eqref{e:cond1} holds and at least there is one equality in \eqref{e:lb1} or \eqref{e:ub} does not hold, then for each $j\not\in S$, it holds
\begin{equation*}
|\langle\A\e_j,\A\x\rangle|<\max_{i\in S}|\langle\A\e_i, \A\x\rangle|.
\end{equation*}
Therefore, it suffices to show that the equality in \eqref{e:lb1} and the equation in \eqref{e:ub} can not hold simultaneously.

Suppose both the equality in \eqref{e:lb1} and the equation in \eqref{e:ub} hold, then by Lemma \ref{l:main1}, $\parallel\A\e_j\parallel_2^2=1+\delta_{K+1}$. Let $\C=(\A_{S\bigcup j})^T\A_{S\bigcup j}$, then $C_{jj}=\parallel\A\e_j\parallel_2^2=1+\delta_{K+1}$, thus for each $i\in S$,
$C_{ij}=0$. In fact, suppose there exists one $i\in S$ such that $C_{ij}\neq0$, then
\begin{equation*}
\parallel\A_{S\bigcup j}^T\A_{S\bigcup j}\e_j\parallel_2\geq \sqrt{\sum_{i\in S}C_{ij}^2}>1+\delta_{K+1}
\end{equation*}
which contradicts Lemma \ref{l:main2}. Therefore, for each $i\in S$,
$C_{ij}=0$. However, in this case, we have
\begin{equation*}
|\langle\A\e_j,\A\x\rangle|=0
\end{equation*}
which contradicts the equality in \eqref{e:ub}. Thus the equality in \eqref{e:lb1} and the equation in \eqref{e:ub} can not hold simultaneously. Therefore,
if \eqref{e:cond1} holds, then the OMP can choose a correct index in the first iteration.

By applying the method used in \cite{MoS12} or \cite{WanS12} and the aforementioned proof, one can similarly show that if \eqref{e:cond1} holds, then the OMP can  choose a correct index in the latter iterations, so the theorem is proved.


\section{ Future Work}

In the future, we will prove or disprove whether $\frac{1}{\sqrt{K}+1}<\delta_{K+1}<\frac{1}{\sqrt{K+1}}$ is a sufficient condition for the OMP to recover every $K-$spares signal $\x$ from $\y=\A\x$ in $K$ iterations.

\vskip3pt
\ack{This work has been supported by NSFC (Grant No. 11201161, 11171125, 91130003) and FRQNT.
}

\vskip5pt

\noindent Jinming Wen (\textit{Dept. of Mathematics and Statistics, McGill University, Canada, H3A 2K6})
\vskip3pt

\noindent Xiaomei Zhu (\textit{College of Electronics and Information Engineering, Nanjing University of Technology,  China, 211816; Dept. of Electrical and Computer Engineering,  McGill University, Canada, H3A 2A7})

\vskip3pt
\noindent E-mail: njiczxm@njut.edu.cn

\noindent Dongfang Li (\textit{School of Mathematics and Statistics, Huazhong University of Science and
Technology, China, 430074; Dept. of Mathematics and Statistics, McGill University, Canada, H3A 2K6})

\end{document}